\documentstyle[epsf]{elsart}

\def\beq{\begin{equation}}
\def\eeq{\end{equation}}
\def\beqar{\begin{eqnarray}}
\def\eeqar{\end{eqnarray}}
\def\barr#1{\begin{array}{#1}}
\def\earr{\end{array}}
\def\bfi{\begin{figure}}
\def\efi{\end{figure}}
\def\btab{\begin{table}}
\def\etab{\end{table}}
\def\bce{\begin{center}}
\def\ece{\end{center}}
\def\nn{\nonumber}

\def\text{\textstyle}

\def\al{\alpha}

\def\de{\delta}

\def\De{\Delta}

\def\refeq#1{\mbox{(\ref{#1})}}
\def\reffi#1{\mbox{Fig.~\ref{#1}}}

\def\citere#1{\mbox{Ref.~\cite{#1}}}
\def\citeres#1{\mbox{Refs.~\cite{#1}}}

\def\solid{\raise.9mm\hbox{\protect\rule{1.1cm}{.2mm}}}
\def\dash{\raise.9mm\hbox{\protect\rule{2mm}{.2mm}}\hspace*{1mm}}

\newcommand{\GeV}{\unskip\,\mathrm{GeV}}

\newcommand{\TeV}{\unskip\,\mathrm{TeV}}

\def\mathswitchr#1{\relax\ifmmode{\mathrm{#1}}\else$\mathrm{#1}$\fi}

\newcommand{\PW}{\mathswitchr W}
\newcommand{\PZ}{\mathswitchr Z}

\newcommand{\PH}{\mathswitchr H}

\newcommand{\Pt}{\mathswitchr t}

\def\mathswitch#1{\relax\ifmmode#1\else$#1$\fi}

\newcommand{\MW}{\mathswitch {M_\PW}}

\newcommand{\MZ}{\mathswitch {M_\PZ}}
\newcommand{\MH}{\mathswitch {M_\PH}}

\newcommand{\Mt}{\mathswitch {m_\Pt}}

\newcommand{\scrs}{\scriptscriptstyle}

\newcommand{\GF}{\mathswitch {G_\mu}}

\newcommand{\mathem}{{\em Mathematica}}
\newcommand{\se}{self-en\-er\-gy}
\newcommand{\ses}{self-en\-er\-gies}
\newcommand{\fea}{{\em FeynArts}}
\newcommand{\fec}{{\em FeynCalc}}
\newcommand{\two}{{\em TwoCalc}}

\begin{document}
\null
\hfill KA-TP-25-1996\\
\null
\hfill WUE-ITP-96.030\\
\null
\hfill hep-ph/9611445\\
\vskip .8cm
\begin{center}
{\Large \bf Calculation of two-loop top-quark\\[.5em]
and Higgs-boson corrections\\[.5em]
in the electroweak Standard Model}
\vskip 2.5em
{\large 
{\sc Stefan Bauberger}%
\footnote{Work supported by the German Federal Ministry of Education,
Science, Research and Technology (BMBF) under contract number 
05 7WZ91P (0).}\\[1ex]
{\normalsize \it
Institut f\"ur Theoretische Physik, Universit\"at W\"urzburg, Am
Hubland, D-97074~W\"urzburg, Germany}\\[2ex]
{\sc Georg Weiglein}\\[1ex]
{\normalsize \it Institut f\"ur Theoretische Physik, Universit\"at Karlsruhe,
D-76128 Karlsruhe, Germany}
}
\vskip 2em
\end{center} \par
\vskip 1.2cm
\vfil
{\bf Abstract} \par
A combination of algebraical and numerical techniques
for calculating two-loop top-quark and Higgs-boson corrections
to electroweak precision observables like $\Delta r$ or the 
$\rho$-parameter is presented. The renormalization is performed within
the on-shell scheme. The results of the calculations are valid for
arbitrary values of $\Mt$, $\MH$ and of the gauge-boson masses.
An example is treated where the full result is compared to the 
result obtained via an expansion up to next-to-leading order in $\Mt$.
As an application, results for the Higgs-mass dependent
top-contributions to $\De r$ are given.
\par
\vskip 1cm
\null
\setcounter{page}{0}
\clearpage

\begin{frontmatter}
\title{Calculation of Two-loop Top-quark and Higgs-boson
Corrections in the Electroweak Standard Model}
\vspace{-0.5cm}
\author[Wuerzburg]{S. Bauberger\thanksref{BMBF}}
and
\author[Karlsruhe]{G. Weiglein}
\address[Wuerzburg]{Institut f\"ur Theoretische Physik,
Universit\"at W\"urzburg, Am Hubland, D-97074~W\"urzburg, Germany}
\address[Karlsruhe]{Institut f\"ur Theoretische Physik,
Universit\"at Karlsruhe, Postfach 6980, D-76128~Karlsruhe, Germany}
\thanks[BMBF]{Work supported by the German Federal Ministry of Education,
Science,
Research and Technology (BMBF) under contract number
05 7WZ91P (0).}
\vspace{-10pt}
\begin{abstract}
A combination of algebraical and numerical techniques
for calculating two-loop top-quark and Higgs-boson corrections
to electroweak precision observables like $\Delta r$ or the 
$\rho$-parameter is presented. The renormalization is performed within
the on-shell scheme. The results of the calculations are valid for
arbitrary values of $\Mt$, $\MH$ and of the gauge-boson masses.
An example is treated where the full result is compared to the 
result obtained via an expansion up to next-to-leading order in $\Mt$.
As an application, results for the Higgs-mass dependent
top-contributions to $\De r$ are given.
\end{abstract}
\end{frontmatter}

%\thispagestyle{empty}
%\setcounter{page}{1}
 
%%\copyrightheading{}                     %{Vol. 0, No. 0 (1993) 000--000}
 
%\vspace*{0.88truein}
 
%%\publisher{(received date)}{(revised date)}
 
%%\vspace*{0.21truein}

%%\vspace*{10pt}
%%\keywords{The contents of the keywords}
 
%%\textlineskip                  %) USE THIS MEASUREMENT WHEN THERE IS
%%\vspace*{12pt}                 %) NO SECTION HEADING
 
%%\vspace*{1pt}\textlineskip      %) USE THIS MEASUREMENT WHEN THERE IS

\vspace{-30pt}
\section{Introduction}          %) A SECTION HEADING
%\vspace*{-0.5pt}
\vspace{-20pt}
\noindent
After the discovery of the top-quark %~\cite{mtexp}
the Higgs-boson remains the only
missing ingredient of the (minimal) electroweak Standard Model (SM).
At the moment, the mass of the Higgs-boson, $\MH$, can only very mildly
be constrained by confronting the SM with precision data.
In order to improve on this situation, a further reduction of the
experimental and theoretical errors is necessary. 
Concerning the reduction of the theoretical error due to missing higher
order corrections, in particular two-loop top-quark corrections and
higher order QCD corrections to the quantity $\Delta r$ derived from
muon decay and to the $\rho$-parameter are of interest.
Recently the leading two-loop top-quark and Higgs-boson corrections to these
quantities~\cite{vdBH} have been supplemented by the calculation
of the 
full Higgs-boson dependence of the leading $\Mt^4$
contribution~\cite{barb2} and furthermore by inclusion of the
next-to-leading top-quark contributions~\cite{gamb}.
Since both the Higgs-mass dependence of the leading $\Mt^4$ contribution
and the inclusion of the next-to-leading term in the $\Mt$~expansion
turned out to yield sizable contributions, a more complete
calculation of the top-quark corrections would be desirable, where no
expansion in $\Mt$ or $\MH$ is made.

\vspace{-8pt}
In this article we describe techniques with which such a calculation
can be carried out. The calculation of top-quark
contributions to $\Delta r$ and other processes with light external
fermions at low energies requires in particular
the evaluation of two-loop
self-energies at non-zero external momentum, while vertex and box
contributions can mostly be reduced to vacuum integrals. We perform the
renormalization within the complete on-shell scheme, i.e.\ we use
physical parameters throughout. 
All calculations are performed in an arbitrary
$R_{\xi}$~gauge with free gauge parameters $\xi_i$, $i = A, Z, W$.

\vspace{-30pt}
\section{Algebraic Evaluation}
\vspace{-20pt}
%\label{sect:Two}
\noindent
The problem of evaluating top-quark or Higgs-boson
corrections to four-fermion processes with light external
particles at low energies in the on-shell renormalization
scheme requires the calculation of two-loop self-energies
with massive particles at non-zero external momentum. The algebraic
evaluation of these contributions can be carried out in a
highly automatized way by means of the \mathem\ packages
\fea~2.0~\cite{fea} and \two~\cite{two}.

\vspace{-8pt}
The package \fea~2.0 is used for generating the relevant one-loop and
two-loop diagrams and counterterm graphs.
For this purpose, we have implemented the complete one-loop
counterterm contributions of the SM (using the complete on-shell
renormalization scheme) as well as the appropriate two-loop terms 
into the model file of the package. 

\begin{figure}
\begin{picture}(350,30)
\epsfxsize=14cm
\put(-32,-30){\epsffile[100 600 600 750]{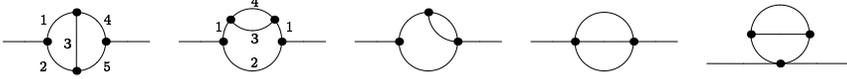}}
\end{picture}
\caption{One-particle irreducible topologies of two-loop \ses\
which do not factorize into one-loop contributions.}
\label{fig:top}
\end{figure}

\vspace{-8pt}
The package \two\ reduces general two-loop self-energies to a minimal
set of standard scalar two-loop integrals of the form 
\vspace{-20pt}
\beqar
T_{i_1 i_2 \ldots i_l}(p^2;m_1^2,m_2^2,\ldots,m_{l}^2 ) = %} \nn \\
\hspace{7 cm} && \nn \\
\int \frac{d^D q_1}{i \pi ^2 (2 \pi \mu)^{D - 4}}
\int \frac{d^D q_2}{i \pi ^2 (2 \pi \mu)^{D - 4}}
\frac{1}{ \Bigl[k_{i_1}^2 - m_1^2 \Bigr] \Bigl[k_{i_2}^2 -
m_2^2 \Bigr] \cdots \Bigl[k_{i_l}^2 - m_l^2 \Bigr] }, && % }
\label{eq:Tint}
\eeqar
which correspond to the basic two-loop topologies depicted in
\reffi{fig:top}. 
In \refeq{eq:Tint} $k_{i_l}$ denotes the momentum of the $i$-th propagator
and $m_l$ its mass. The $k_{i_l}$ are related to the integration momenta
$q_1$ and $q_2$ and the external momentum $p$ via
$k_1 = q_1, \; k_2 = q_1 + p, \; k_3 = q_2 - q_1, \; k_4 = q_2,
\; k_5 = q_2 + p$. %\label{eq:momenta}
The topologies shown in Fig.~\ref{fig:top} correspond to the
scalar integrals $T_{12345}$, $T_{11234}$, $T_{1234}$, $T_{234}$, and
$T_{1134}$, respectively. The analytical expression for
$T_{11234}$ can be obtained from $T_{1234}$ by
partial fractioning or taking the
derivative with respect to $m_1^2$.
Other integrals with higher powers of propagators are treated
in the same way.
For the general case one therefore needs to evaluate only
four different types of two-loop scalar integrals.

\vspace{-8pt}
The algebraic evaluation with \two\ is performed for arbitrary values of
all particle masses, the invariant momentum $p^2$ and the space-time
dimension $D$. A general $R_{\xi}$~gauge is used which is specified by
one gauge parameter $\xi_i$, $i = A, Z, W$, for each vector boson.
The algorithm of \two\ has been described in detail in \citeres{two}.
For the contraction of Lorentz indices, reduction of the Dirac algebra
and evaluation of Dirac traces, which can be worked out like in the
one-loop case, the routines of the program \fec~\cite{fec} are used.
The further evaluation is based on a method for the tensor integral
decomposition of two-loop \ses~\cite{two} and furthermore makes use of
certain symmetry properties of the two-loop integrals.

\vspace{-8pt}
The counterterm contributions are also calculated with \two.
The sum of unrenormalized two-loop diagrams and diagrams with
counterterm insertions is expressed in terms of a minimal basis of
standard scalar integrals. This is a very convenient feature, 
since it allows to directly check at the algebraic level whether the 
gauge-parameter dependence of the result drops out. As a further nontrivial 
check which can directly be read off from the algebraic result, all
field renormalization constants of internal particles must drop out in
the sum of all contributing diagrams. After inserting the divergent part
of the two-loop integrals and decomposing the one-loop integrals into
divergent and finite parts, also the UV-finiteness of the result can be
checked algebraically, i.e.\ all terms proportional to $1/\de^2$ and
$1/\de$, where $\de = (4 - D)/2$, must cancel.

\vspace{-30pt}
\section{Two-loop On-shell Renormalization}
\vspace{-20pt}
\label{sect:ren}
\noindent
For our calculations we use the complete on-shell renormalization scheme
at the two-loop level, i.e.\ we use as parameters the masses of the
physical particles and the electric charge $e$, and we introduce field
renormalizations such that the residues of all propagators are
equal to one (see e.g.~\citere{Dehab}). 
The formulation in terms of physical parameters is
particularly important in view of calculating corrections to the
quantity $\De r$ derived from muon decay, which relates the Fermi
constant $\GF$ to the W-boson and Z-boson masses, $\MW$ and $\MZ$,
and to the electromagnetic fine structure constant $\al = e^2/(4 \pi)$.

\vspace{-8pt}
We perform the renormalization in such a way that the gauge-fixing 
term in the Lagrangian does not give rise
to counterterm contributions, i.e.\ the 
renormalization of the parameters and fields in the gauge-fixing term is
canceled by appropriate renormalizations of the gauge parameters.
It should be noted
that the renormalization prescription chosen for the gauge-fixing
term determines the renormalization of the ghost sector, which at the
two-loop level also enters the quantities in the physical sector of the
theory. According to the described renormalization procedure
we have derived the counterterms for the two-point and
three-point vertices containing ghost fields. For the example of the
complete two-loop top-quark contribution to the W-boson \se\ we have
checked that insertion of these renormalization constants does in fact
yield a finite result for the two-loop \se.

\vspace{-8pt}
As a specific example we briefly consider the 
Higgs-dependent top-quark corrections to $\Delta r$.
It is easy to see that Higgs-dependent top-quark corrections can, 
except for the renormalization, enter $\De r$ only via two-point
functions. 
Besides these \se\ corrections
one also has to consider contributions arising from the
renormalization of the relevant one-loop and two-loop three-point
functions. The two-loop renormalization constants receiving
Higgs-dependent top-quark corrections are 
the charge renormalization constant, $\de Z_{e, (2)}$, and
the counterterm of the electroweak mixing angle, 
$\de s^2_{{\scrs\PW}, (2)}$,
while the field renormalization constant 
$\de Z_{W, (2)}$ (and also $\de Z_{W, (1)}$) cancels in the
sum of the contributing diagrams.

\vspace{-30pt}
\section{Decomposition and numerical Evaluation of scalar %Two-Loop 
Integrals}
\vspace{-20pt}
\label{sect:decomp}
\noindent
It is known since several years that massive two-loop integrals
are in general not expressible in terms of polylogarithmic 
functions~\cite{ScharfDipl}.
Our approach for a numerical evaluation of scalar self-energy integrals are
one-dimensional integral representations.
They allow
for a very fast calculation of these functions with high precision.

\vspace{-8pt}
Such a numerical approach can of course 
only be applied to finite functions,
while UV- and IR-divergencies of the two-loop integrals
have to be kept under control. We decompose the integrals into divergent
and finite parts and check algebraically  that
the divergencies cancel. For some integrals we make use of the fact
that the two-loop vacuum integrals can be calculated analytically
(see e.g.\ \citere{DavyTausk}).
Then we numerically calculate suitable combinations of
the non-vacuum and the vacuum integrals.
For some diagrams it proved
useful to separate the two-particle cut and the three-particle cut
contributions~\cite{BBBB}, or to combine both methods.
As an example the
momentum derivative of $T_{11234}$ can be calculated as
\vspace{-20pt}
\begin{eqnarray}
 \!\!\!\!
  \frac{\partial}{\partial p^2}  T_{11234}(p^2,m_1^2,m_2^2,m_3^2,m_4^2)
 &=&
   \frac{\partial}{\partial m_1^2}\left( B_0(m_1^2,m_3^2,m_4^2)
     \frac{\partial}{\partial p^2} B_0(p^2,m_1^2,m_2^2) \right)
 \nonumber \\
    &&+
    DT_{11234C3}(p^2,m_1^2,m_2^2,m_3^2,m_4^2) \,,
         \label{defDT11234C3A}
\end{eqnarray}
where $DT_{11234C3}$, the three-particle cut contribution, is finite,
while the divergencies are contained in one-loop $B_0$-integrals, which 
can be calculated analytically.

\vspace{-8pt}
For the numerical evaluation
there remains an expression for the finite part which
is in general very extensive.
The main ingredients of the numerical evaluation
are one-dimensional integral representations which we have implemented
in C++. For the integration we apply an adaptive Gauss-Kronrod 
algorithm~\cite{QUADPACK}.

\vspace{-8pt}
For those topologies which contain a one-loop self-energy insertion
we make use of a dispersion representation of the subloop which has been
discussed in detail in \citeres{BBBB}.
In these representations the two-particle cut contributions
have been separated as  products of $B_0$-functions. All momentum or mass
derivatives of the functions $T_{234}$ and
$T_{1234}$ can be calculated starting from these formulae.
For the master topology, $T_{12345}$, we refer to the formulae presented
in \citere{BB}.

\vspace{-8pt}
The basic functions in our integration kernels are $B_0$-functions and
the discontinuities of $B_0$-functions. In some cases it is useful
for the numerical stability
to subtract the asymptotic behavior of the $B_0$-function for a
large mass-variable. It is given by
\vspace{-10pt}
\begin{equation}
 \!\!\!\!\!\!\!\!\!\!\!\! B_0(p^2;s,m^2) = \frac{1}{\delta} + 1 
                       - \log\frac{s}{4 \pi \mu^2}
                       + \frac{m^2}{s} \log\frac{m^2}{s}
                       + \frac{p^2}{2\,s}
                       + B_{0rest}(p^2;s,m^2) \,.
\end{equation}
\vspace{-10pt}

\vspace{-8pt}
Calculating physical processes
we encountered in some
parameter regions huge cancellations among the contributions
of the diagrams or the scalar integrals. Therefore we perform
in these cases the calculations inside the integral representations with
quadruple precision~\cite{Briggs}.
Typical computation times are $0.03$ seconds for the evaluation of the
functions $T_{234}$ and $T_{1234}$
and $0.3$ seconds for the evaluation of $T_{12345}$
to ten digits precision on a
workstation DEC 3000 AXP. Using quadruple precision slows down
the calculations by a factor of about $10$, which results e.g.
in $40$ seconds for the calculation of the 
Higgs-dependent top-quark contributions to the W-boson \se\
for one set of parameters.

\setcounter{footnote}{0}
\vspace{-30pt}
\section{Results}
\vspace{-20pt}
\label{sect:res}
\subsection{Comparison with the expansion in $\Mt$}
\begin{figure}[htb]
\begin{picture}(350,40)
\put(72,-283){\includegraphics{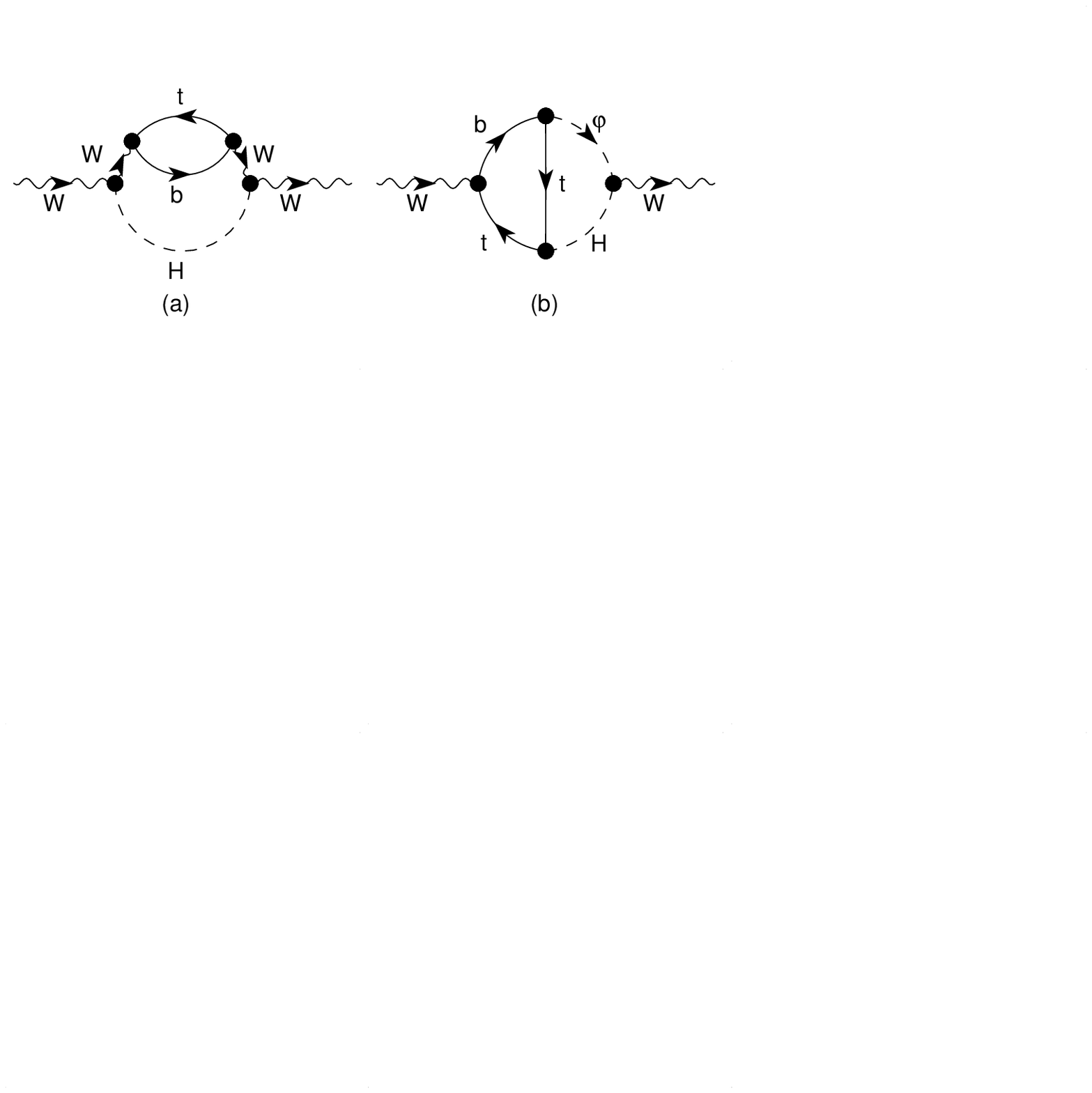}}
\end{picture}
\caption{The diagrams for which we compare our results
with the expansion in~$\Mt$.} 
\label{fig:diags}
\end{figure}

\vspace{-20pt}
\noindent
In this section we compare our results for the two diagrams shown
in \reffi{fig:diags}
with the results of an expansion up to next-to-leading order in
$\Mt$\footnote{P.~Gambino has kindly provided us with his results for
the expansions of these diagrams.} (see \citere{gamb}), which
takes into account terms of order $\Mt^4$ and $\Mt^2$. This 
expansion is performed in two regions, namely in
the light Higgs region (``light Higgs expansion'') and in the 
heavy Higgs region (``heavy Higgs expansion'')~\cite{gamb}. 

\begin{figure}[htb]
\hspace{1.8cm}
% GNUPLOT: LaTeX picture with Postscript
\setlength{\unitlength}{0.1bp}
\special{!
%!PS-Adobe-2.0
%%Creator: gnuplot
%%DocumentFonts: Helvetica
%%BoundingBox: 50 50 626 418
%%Pages: (atend)
%%EndComments
/gnudict 40 dict def
gnudict begin
/Color false def
/Solid false def
/gnulinewidth 5.000 def
/vshift -33 def
/dl {10 mul} def
/hpt 31.5 def
/vpt 31.5 def
/M {moveto} bind def
/L {lineto} bind def
/R {rmoveto} bind def
/V {rlineto} bind def
/vpt2 vpt 2 mul def
/hpt2 hpt 2 mul def
/Lshow { currentpoint stroke M
  0 vshift R show } def
/Rshow { currentpoint stroke M
  dup stringwidth pop neg vshift R show } def
/Cshow { currentpoint stroke M
  dup stringwidth pop -2 div vshift R show } def
/DL { Color {setrgbcolor Solid {pop []} if 0 setdash }
 {pop pop pop Solid {pop []} if 0 setdash} ifelse } def
/BL { stroke gnulinewidth 2 mul setlinewidth } def
/AL { stroke gnulinewidth 2 div setlinewidth } def
/PL { stroke gnulinewidth setlinewidth } def
/LTb { BL [] 0 0 0 DL } def
/LTa { AL [1 dl 2 dl] 0 setdash 0 0 0 setrgbcolor } def
/LT0 { PL [] 0 1 0 DL } def
/LT1 { PL [4 dl 2 dl] 0 0 1 DL } def
/LT2 { PL [2 dl 3 dl] 1 0 0 DL } def
/LT3 { PL [1 dl 1.5 dl] 1 0 1 DL } def
/LT4 { PL [5 dl 2 dl 1 dl 2 dl] 0 1 1 DL } def
/LT5 { PL [4 dl 3 dl 1 dl 3 dl] 1 1 0 DL } def
/LT6 { PL [2 dl 2 dl 2 dl 4 dl] 0 0 0 DL } def
/LT7 { PL [2 dl 2 dl 2 dl 2 dl 2 dl 4 dl] 1 0.3 0 DL } def
/LT8 { PL [2 dl 2 dl 2 dl 2 dl 2 dl 2 dl 2 dl 4 dl] 0.5 0.5 0.5 DL } def
/P { stroke [] 0 setdash
  currentlinewidth 2 div sub M
  0 currentlinewidth V stroke } def
/D { stroke [] 0 setdash 2 copy vpt add M
  hpt neg vpt neg V hpt vpt neg V
  hpt vpt V hpt neg vpt V closepath stroke
  P } def
/A { stroke [] 0 setdash vpt sub M 0 vpt2 V
  currentpoint stroke M
  hpt neg vpt neg R hpt2 0 V stroke
  } def
/B { stroke [] 0 setdash 2 copy exch hpt sub exch vpt add M
  0 vpt2 neg V hpt2 0 V 0 vpt2 V
  hpt2 neg 0 V closepath stroke
  P } def
/C { stroke [] 0 setdash exch hpt sub exch vpt add M
  hpt2 vpt2 neg V currentpoint stroke M
  hpt2 neg 0 R hpt2 vpt2 V stroke } def
/T { stroke [] 0 setdash 2 copy vpt 1.12 mul add M
  hpt neg vpt -1.62 mul V
  hpt 2 mul 0 V
  hpt neg vpt 1.62 mul V closepath stroke
  P  } def
/S { 2 copy A C} def
end
}
\begin{picture}(2880,1576)(0,0)
\special{"
gnudict begin
gsave
50 50 translate
0.100 0.100 scale
0 setgray
/Helvetica findfont 100 scalefont setfont
newpath
-500.000000 -500.000000 translate
LTa
600 1047 M
2097 0 V
600 251 M
0 1274 V
LTb
600 251 M
63 0 V
2034 0 R
-63 0 V
600 410 M
63 0 V
2034 0 R
-63 0 V
600 570 M
63 0 V
2034 0 R
-63 0 V
600 729 M
63 0 V
2034 0 R
-63 0 V
600 888 M
63 0 V
2034 0 R
-63 0 V
600 1047 M
63 0 V
2034 0 R
-63 0 V
600 1207 M
63 0 V
2034 0 R
-63 0 V
600 1366 M
63 0 V
2034 0 R
-63 0 V
600 1525 M
63 0 V
2034 0 R
-63 0 V
600 251 M
0 63 V
0 1211 R
0 -63 V
950 251 M
0 63 V
0 1211 R
0 -63 V
1299 251 M
0 63 V
0 1211 R
0 -63 V
1649 251 M
0 63 V
0 1211 R
0 -63 V
1998 251 M
0 63 V
0 1211 R
0 -63 V
2348 251 M
0 63 V
0 1211 R
0 -63 V
2697 251 M
0 63 V
0 1211 R
0 -63 V
600 251 M
2097 0 V
0 1274 V
-2097 0 V
600 251 L
LT0
2408 630 M
180 0 V
706 1111 M
10 51 V
11 41 V
10 34 V
10 28 V
11 23 V
10 20 V
11 17 V
10 14 V
11 12 V
10 11 V
11 9 V
10 8 V
11 6 V
10 6 V
11 5 V
10 4 V
11 4 V
10 3 V
11 3 V
41 7 V
70 4 V
70 -2 V
70 -5 V
70 -6 V
70 -7 V
69 -8 V
70 -7 V
70 -7 V
70 -7 V
70 -7 V
70 -7 V
70 -6 V
70 -6 V
70 -6 V
70 -5 V
69 -6 V
70 -5 V
70 -4 V
70 -5 V
70 -5 V
70 -4 V
70 -4 V
70 -4 V
70 -4 V
70 -5 V
LT1
2408 500 M
180 0 V
706 722 M
10 34 V
11 27 V
10 23 V
10 19 V
11 16 V
10 13 V
11 12 V
10 10 V
11 10 V
10 8 V
11 7 V
10 7 V
11 6 V
10 5 V
11 5 V
10 5 V
11 4 V
10 4 V
11 4 V
41 12 V
70 15 V
70 11 V
70 9 V
70 6 V
70 5 V
69 5 V
70 3 V
70 3 V
70 3 V
70 2 V
70 2 V
70 2 V
70 1 V
70 2 V
70 1 V
69 1 V
70 2 V
70 1 V
70 1 V
70 1 V
70 0 V
70 2 V
70 0 V
70 0 V
70 0 V
LT2
2408 370 M
180 0 V
706 369 M
10 25 V
11 20 V
10 17 V
10 15 V
11 14 V
10 12 V
11 11 V
10 10 V
11 9 V
10 9 V
11 8 V
10 8 V
11 7 V
10 7 V
11 7 V
10 7 V
11 6 V
10 6 V
11 5 V
41 21 V
70 30 V
70 25 V
70 21 V
70 19 V
70 16 V
69 15 V
70 13 V
70 11 V
70 11 V
70 10 V
70 9 V
70 8 V
70 8 V
70 7 V
70 6 V
69 7 V
70 5 V
70 6 V
70 5 V
70 5 V
70 5 V
70 4 V
70 4 V
70 4 V
70 4 V
stroke
grestore
end
showpage
}
\put(2348,370){\makebox(0,0)[r]{$\MH/\Mt={350}/{175}$}}
\put(2348,500){\makebox(0,0)[r]{$\MH/\Mt={300}/{175}$}}
\put(2348,630){\makebox(0,0)[r]{$\MH/\Mt={250}/{175}$}}
\put(2968,151){\makebox(0,0){$\frac{\Mt}{\GeV}$}}
\put(-80,988){%
%\special{ps: gsave currentpoint currentpoint translate
%270 rotate neg exch neg exch translate}%
\makebox(0,0)[b]{\shortstack{$\frac{Re(\Delta \Sigma)}{\MW^2 \Mt}/ \frac{\alpha^2}{4 \pi^2}/\GeV^{-1}$}}%
%\special{ps: currentpoint grestore moveto}%
}
\put(2697,151){\makebox(0,0){3000}}
\put(2348,151){\makebox(0,0){2500}}
\put(1998,151){\makebox(0,0){2000}}
\put(1649,151){\makebox(0,0){1500}}
\put(1299,151){\makebox(0,0){1000}}
\put(950,151){\makebox(0,0){500}}
\put(600,151){\makebox(0,0){0}}
\put(540,1525){\makebox(0,0)[r]{0.06}}
\put(540,1366){\makebox(0,0)[r]{0.04}}
\put(540,1207){\makebox(0,0)[r]{0.02}}
\put(540,1047){\makebox(0,0)[r]{0}}
\put(540,888){\makebox(0,0)[r]{-0.02}}
\put(540,729){\makebox(0,0)[r]{-0.04}}
\put(540,570){\makebox(0,0)[r]{-0.06}}
\put(540,410){\makebox(0,0)[r]{-0.08}}
\put(540,251){\makebox(0,0)[r]{-0.1}}
\end{picture}
\vspace{-0.6cm}
\caption{
Diagram (a): Difference between full calculation and heavy Higgs expansion,
divided by $\Mt$ ($p^2=\MW^2$).
}
\label{Gambino-expansion-test}
\end{figure}
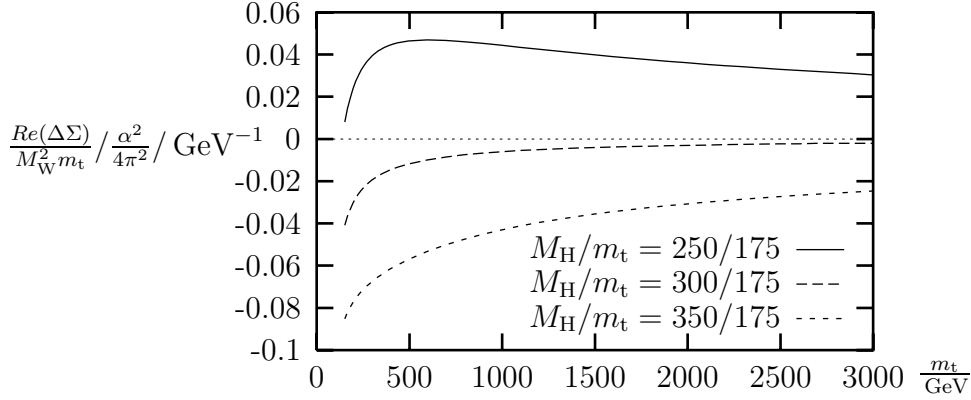

\vspace{-8pt}
We first consider asymptotically large values of $\Mt$ and check whether
our full result and the expansion in $\Mt$ agree in this region.
In order to compare with the heavy Higgs expansion, both $\Mt$ and $\MH$
are made asymptotically large, while for comparison with the light Higgs
expansion $\MH$ is kept fixed. \reffi{Gambino-expansion-test} shows 
for diagram (a) the difference between the finite parts of our result 
and the heavy Higgs expansion, divided by $\Mt$, as a function of $\Mt$.
For the self-energies we have here and below used units of
$\MW^2\frac{\alpha^2}{4 \pi^2}$, which amounts with
$\alpha(\MZ^2)=\frac{1}{128.9}$ to
$1.52 \times 10^{-6} \MW^2$. The difference between the full result
and the expansions in $\Mt$ is expected to be at most 
of the order $\Mt$, as it is the case in \reffi{Gambino-expansion-test}.
We have checked that also for the light Higgs expansion of diagram (a)
and for both expansions of diagram (b) one finds agreement between the
full result and the expansions in the asymptotic region. 

\vspace{-8pt}
Next we consider the physical parameter 
region, i.e.\ we put $\Mt=175 \GeV$,
and compare the full result with the expansions for different values
of $\MH$. This is shown in \reffi{Gamb-a-mh} for diagram (a) and in
\reffi{Gamb-b-mh} for diagram (b). We find relatively good agreement
for diagram (b), while for diagram (a) for most values of $\MH$ the
relative deviations are larger. It should be noted, however, that the
numerical contribution of diagram (b) is very large so that in the final
result large cancellations can be expected.

\begin{figure}[htb]
\hspace{1.8cm}
% GNUPLOT: LaTeX picture with Postscript
\setlength{\unitlength}{0.1bp}
\special{!
%!PS-Adobe-2.0
%%Creator: gnuplot
%%DocumentFonts: Helvetica
%%BoundingBox: 50 50 626 418
%%Pages: (atend)
%%EndComments
/gnudict 40 dict def
gnudict begin
/Color false def
/Solid false def
/gnulinewidth 5.000 def
/vshift -33 def
/dl {10 mul} def
/hpt 31.5 def
/vpt 31.5 def
/M {moveto} bind def
/L {lineto} bind def
/R {rmoveto} bind def
/V {rlineto} bind def
/vpt2 vpt 2 mul def
/hpt2 hpt 2 mul def
/Lshow { currentpoint stroke M
  0 vshift R show } def
/Rshow { currentpoint stroke M
  dup stringwidth pop neg vshift R show } def
/Cshow { currentpoint stroke M
  dup stringwidth pop -2 div vshift R show } def
/DL { Color {setrgbcolor Solid {pop []} if 0 setdash }
 {pop pop pop Solid {pop []} if 0 setdash} ifelse } def
/BL { stroke gnulinewidth 2 mul setlinewidth } def
/AL { stroke gnulinewidth 2 div setlinewidth } def
/PL { stroke gnulinewidth setlinewidth } def
/LTb { BL [] 0 0 0 DL } def
/LTa { AL [1 dl 2 dl] 0 setdash 0 0 0 setrgbcolor } def
/LT0 { PL [] 0 1 0 DL } def
/LT1 { PL [4 dl 2 dl] 0 0 1 DL } def
/LT2 { PL [2 dl 3 dl] 1 0 0 DL } def
/LT3 { PL [1 dl 1.5 dl] 1 0 1 DL } def
/LT4 { PL [5 dl 2 dl 1 dl 2 dl] 0 1 1 DL } def
/LT5 { PL [4 dl 3 dl 1 dl 3 dl] 1 1 0 DL } def
/LT6 { PL [2 dl 2 dl 2 dl 4 dl] 0 0 0 DL } def
/LT7 { PL [2 dl 2 dl 2 dl 2 dl 2 dl 4 dl] 1 0.3 0 DL } def
/LT8 { PL [2 dl 2 dl 2 dl 2 dl 2 dl 2 dl 2 dl 4 dl] 0.5 0.5 0.5 DL } def
/P { stroke [] 0 setdash
  currentlinewidth 2 div sub M
  0 currentlinewidth V stroke } def
/D { stroke [] 0 setdash 2 copy vpt add M
  hpt neg vpt neg V hpt vpt neg V
  hpt vpt V hpt neg vpt V closepath stroke
  P } def
/A { stroke [] 0 setdash vpt sub M 0 vpt2 V
  currentpoint stroke M
  hpt neg vpt neg R hpt2 0 V stroke
  } def
/B { stroke [] 0 setdash 2 copy exch hpt sub exch vpt add M
  0 vpt2 neg V hpt2 0 V 0 vpt2 V
  hpt2 neg 0 V closepath stroke
  P } def
/C { stroke [] 0 setdash exch hpt sub exch vpt add M
  hpt2 vpt2 neg V currentpoint stroke M
  hpt2 neg 0 R hpt2 vpt2 V stroke } def
/T { stroke [] 0 setdash 2 copy vpt 1.12 mul add M
  hpt neg vpt -1.62 mul V
  hpt 2 mul 0 V
  hpt neg vpt 1.62 mul V closepath stroke
  P  } def
/S { 2 copy A C} def
end
}
\begin{picture}(2880,1576)(0,0)
\special{"
gnudict begin
gsave
50 50 translate
0.100 0.100 scale
0 setgray
/Helvetica findfont 100 scalefont setfont
newpath
-500.000000 -500.000000 translate
LTa
600 830 M
2097 0 V
LTb
600 367 M
63 0 V
2034 0 R
-63 0 V
600 598 M
63 0 V
2034 0 R
-63 0 V
600 830 M
63 0 V
2034 0 R
-63 0 V
600 1062 M
63 0 V
2034 0 R
-63 0 V
600 1293 M
63 0 V
2034 0 R
-63 0 V
600 1525 M
63 0 V
2034 0 R
-63 0 V
755 251 M
0 63 V
0 1211 R
0 -63 V
1144 251 M
0 63 V
0 1211 R
0 -63 V
1532 251 M
0 63 V
0 1211 R
0 -63 V
1920 251 M
0 63 V
0 1211 R
0 -63 V
2309 251 M
0 63 V
0 1211 R
0 -63 V
2697 251 M
0 63 V
0 1211 R
0 -63 V
600 251 M
2097 0 V
0 1274 V
-2097 0 V
600 251 L
LT0
1980 1423 M
180 0 V
654 251 M
25 25 V
67 68 V
57 55 V
50 46 V
45 40 V
42 35 V
39 32 V
37 28 V
34 26 V
33 24 V
32 23 V
30 21 V
29 19 V
28 18 V
28 18 V
26 16 V
26 16 V
24 15 V
25 14 V
23 14 V
24 13 V
22 13 V
22 12 V
22 12 V
21 11 V
21 11 V
20 11 V
21 10 V
19 10 V
20 10 V
19 9 V
19 10 V
18 9 V
18 8 V
18 9 V
18 8 V
18 8 V
17 8 V
98 44 V
152 64 V
139 55 V
128 47 V
120 43 V
114 38 V
107 36 V
103 32 V
25 8 V
LT1
1980 1293 M
180 0 V
600 830 M
79 0 V
67 0 V
57 0 V
50 0 V
45 0 V
42 0 V
39 0 V
37 0 V
34 0 V
33 0 V
32 0 V
30 0 V
29 0 V
28 0 V
28 0 V
26 0 V
26 0 V
24 0 V
25 0 V
23 0 V
24 0 V
22 0 V
22 0 V
22 0 V
21 0 V
21 0 V
20 0 V
21 0 V
19 0 V
20 0 V
19 0 V
19 0 V
18 0 V
18 0 V
18 0 V
18 0 V
18 0 V
17 0 V
98 0 V
152 0 V
139 0 V
128 0 V
120 0 V
114 0 V
107 0 V
103 0 V
25 0 V
LT2
1980 1163 M
180 0 V
600 276 M
79 53 V
67 44 V
57 33 V
50 28 V
45 22 V
42 19 V
39 17 V
37 14 V
34 13 V
33 11 V
32 10 V
30 10 V
29 8 V
28 8 V
28 7 V
26 7 V
26 6 V
24 6 V
25 6 V
23 5 V
24 5 V
22 4 V
22 5 V
22 4 V
21 4 V
21 3 V
20 4 V
21 3 V
19 3 V
20 3 V
19 3 V
19 3 V
18 3 V
18 3 V
18 2 V
18 3 V
18 2 V
17 2 V
98 12 V
152 16 V
139 12 V
128 10 V
120 8 V
114 7 V
107 6 V
103 5 V
25 1 V
stroke
grestore
end
showpage
}
\put(1920,1163){\makebox(0,0)[r]{light Higgs exp.}}
\put(1920,1293){\makebox(0,0)[r]{heavy Higgs exp. ($=0$)}}
\put(1920,1423){\makebox(0,0)[r]{full calculation}}
\put(2968,151){\makebox(0,0){$\frac{\MH}{\GeV}$}}
\put(100,888){%
%\special{ps: gsave currentpoint currentpoint translate
%270 rotate neg exch neg exch translate}%
\makebox(0,0)[b]{\shortstack{$\frac{Re(\Sigma)}{\MW^2}/\frac{\alpha^2}{4 \pi^2}$}}%
%\special{ps: currentpoint grestore moveto}%
}
\put(2697,151){\makebox(0,0){600}}
\put(2309,151){\makebox(0,0){500}}
\put(1920,151){\makebox(0,0){400}}
\put(1532,151){\makebox(0,0){300}}
\put(1144,151){\makebox(0,0){200}}
\put(755,151){\makebox(0,0){100}}
\put(540,1525){\makebox(0,0)[r]{60}}
\put(540,1293){\makebox(0,0)[r]{40}}
\put(540,1062){\makebox(0,0)[r]{20}}
\put(540,830){\makebox(0,0)[r]{0}}
\put(540,598){\makebox(0,0)[r]{-20}}
\put(540,367){\makebox(0,0)[r]{-40}}
\end{picture}
\vspace{-0.6cm}
\caption{
Diagram (a): Full calculation and expansions 
($\Mt=175 \GeV$, $p^2=\MW^2$).}
\label{Gamb-a-mh}
\end{figure}
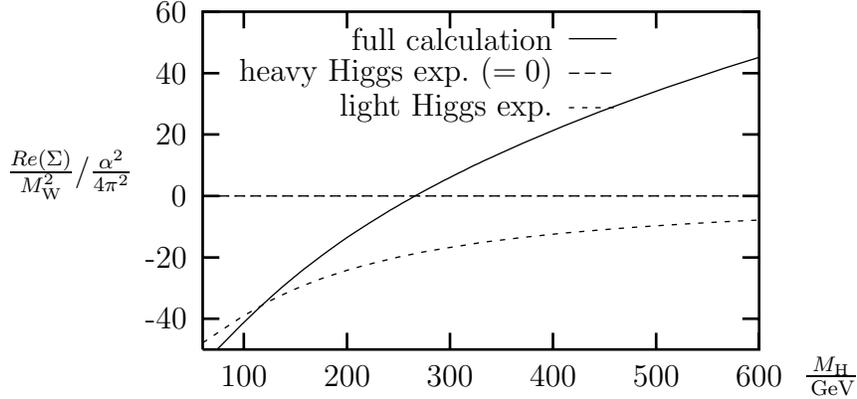
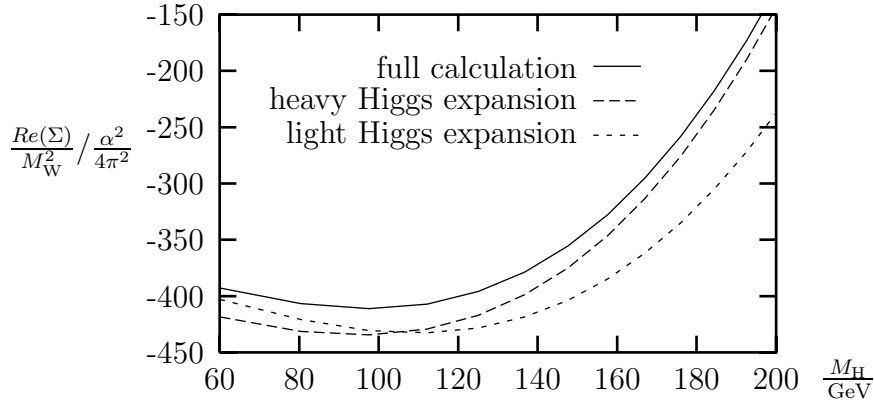
\begin{figure}[htb]
\hspace{1.8cm}
% GNUPLOT: LaTeX picture with Postscript
\setlength{\unitlength}{0.1bp}
\special{!
%!PS-Adobe-2.0
%%Creator: gnuplot
%%DocumentFonts: Helvetica
%%BoundingBox: 50 50 626 418
%%Pages: (atend)
%%EndComments
/gnudict 40 dict def
gnudict begin
/Color false def
/Solid false def
/gnulinewidth 5.000 def
/vshift -33 def
/dl {10 mul} def
/hpt 31.5 def
/vpt 31.5 def
/M {moveto} bind def
/L {lineto} bind def
/R {rmoveto} bind def
/V {rlineto} bind def
/vpt2 vpt 2 mul def
/hpt2 hpt 2 mul def
/Lshow { currentpoint stroke M
  0 vshift R show } def
/Rshow { currentpoint stroke M
  dup stringwidth pop neg vshift R show } def
/Cshow { currentpoint stroke M
  dup stringwidth pop -2 div vshift R show } def
/DL { Color {setrgbcolor Solid {pop []} if 0 setdash }
 {pop pop pop Solid {pop []} if 0 setdash} ifelse } def
/BL { stroke gnulinewidth 2 mul setlinewidth } def
/AL { stroke gnulinewidth 2 div setlinewidth } def
/PL { stroke gnulinewidth setlinewidth } def
/LTb { BL [] 0 0 0 DL } def
/LTa { AL [1 dl 2 dl] 0 setdash 0 0 0 setrgbcolor } def
/LT0 { PL [] 0 1 0 DL } def
/LT1 { PL [4 dl 2 dl] 0 0 1 DL } def
/LT2 { PL [2 dl 3 dl] 1 0 0 DL } def
/LT3 { PL [1 dl 1.5 dl] 1 0 1 DL } def
/LT4 { PL [5 dl 2 dl 1 dl 2 dl] 0 1 1 DL } def
/LT5 { PL [4 dl 3 dl 1 dl 3 dl] 1 1 0 DL } def
/LT6 { PL [2 dl 2 dl 2 dl 4 dl] 0 0 0 DL } def
/LT7 { PL [2 dl 2 dl 2 dl 2 dl 2 dl 4 dl] 1 0.3 0 DL } def
/LT8 { PL [2 dl 2 dl 2 dl 2 dl 2 dl 2 dl 2 dl 4 dl] 0.5 0.5 0.5 DL } def
/P { stroke [] 0 setdash
  currentlinewidth 2 div sub M
  0 currentlinewidth V stroke } def
/D { stroke [] 0 setdash 2 copy vpt add M
  hpt neg vpt neg V hpt vpt neg V
  hpt vpt V hpt neg vpt V closepath stroke
  P } def
/A { stroke [] 0 setdash vpt sub M 0 vpt2 V
  currentpoint stroke M
  hpt neg vpt neg R hpt2 0 V stroke
  } def
/B { stroke [] 0 setdash 2 copy exch hpt sub exch vpt add M
  0 vpt2 neg V hpt2 0 V 0 vpt2 V
  hpt2 neg 0 V closepath stroke
  P } def
/C { stroke [] 0 setdash exch hpt sub exch vpt add M
  hpt2 vpt2 neg V currentpoint stroke M
  hpt2 neg 0 R hpt2 vpt2 V stroke } def
/T { stroke [] 0 setdash 2 copy vpt 1.12 mul add M
  hpt neg vpt -1.62 mul V
  hpt 2 mul 0 V
  hpt neg vpt 1.62 mul V closepath stroke
  P  } def
/S { 2 copy A C} def
end
}
\begin{picture}(2880,1576)(0,0)
\special{"
gnudict begin
gsave
50 50 translate
0.100 0.100 scale
0 setgray
/Helvetica findfont 100 scalefont setfont
newpath
-500.000000 -500.000000 translate
LTa
LTb
600 251 M
63 0 V
2034 0 R
-63 0 V
600 463 M
63 0 V
2034 0 R
-63 0 V
600 676 M
63 0 V
2034 0 R
-63 0 V
600 888 M
63 0 V
2034 0 R
-63 0 V
600 1100 M
63 0 V
2034 0 R
-63 0 V
600 1313 M
63 0 V
2034 0 R
-63 0 V
600 1525 M
63 0 V
2034 0 R
-63 0 V
600 251 M
0 63 V
0 1211 R
0 -63 V
900 251 M
0 63 V
0 1211 R
0 -63 V
1199 251 M
0 63 V
0 1211 R
0 -63 V
1499 251 M
0 63 V
0 1211 R
0 -63 V
1798 251 M
0 63 V
0 1211 R
0 -63 V
2098 251 M
0 63 V
0 1211 R
0 -63 V
2397 251 M
0 63 V
0 1211 R
0 -63 V
2697 251 M
0 63 V
0 1211 R
0 -63 V
600 251 M
2097 0 V
0 1274 V
-2097 0 V
600 251 L
LT0
2008 1330 M
180 0 V
600 494 M
906 435 L
257 -19 V
219 17 V
193 48 V
175 73 V
162 98 V
150 118 V
141 139 V
134 156 V
127 173 V
122 189 V
56 98 V
LT1
2008 1200 M
180 0 V
600 385 M
906 330 L
257 -13 V
219 22 V
193 52 V
175 78 V
162 100 V
150 122 V
141 141 V
134 158 V
127 176 V
122 190 V
96 169 V
LT2
2008 1070 M
180 0 V
600 451 M
906 375 L
257 -42 V
219 -8 V
193 18 V
175 42 V
162 62 V
150 81 V
141 98 V
134 113 V
127 128 V
122 141 V
111 146 V
stroke
grestore
end
showpage
}
\put(1948,1070){\makebox(0,0)[r]{light Higgs expansion}}
\put(1948,1200){\makebox(0,0)[r]{heavy Higgs expansion}}
\put(1948,1330){\makebox(0,0)[r]{full calculation}}
\put(2968,151){\makebox(0,0){$\frac{\MH}{\GeV}$}}
\put(40,988){%
%\special{ps: gsave currentpoint currentpoint translate
%270 rotate neg exch neg exch translate}%
\makebox(0,0)[b]{\shortstack{$\frac{Re(\Sigma)}{\MW^2}/\frac{\alpha^2}{4 \pi^2}$}}%
%\special{ps: currentpoint grestore moveto}%
}
\put(2697,151){\makebox(0,0){200}}
\put(2397,151){\makebox(0,0){180}}
\put(2098,151){\makebox(0,0){160}}
\put(1798,151){\makebox(0,0){140}}
\put(1499,151){\makebox(0,0){120}}
\put(1199,151){\makebox(0,0){100}}
\put(900,151){\makebox(0,0){80}}
\put(600,151){\makebox(0,0){60}}
\put(540,1525){\makebox(0,0)[r]{-150}}
\put(540,1313){\makebox(0,0)[r]{-200}}
\put(540,1100){\makebox(0,0)[r]{-250}}
\put(540,888){\makebox(0,0)[r]{-300}}
\put(540,676){\makebox(0,0)[r]{-350}}
\put(540,463){\makebox(0,0)[r]{-400}}
\put(540,251){\makebox(0,0)[r]{-450}}
\end{picture}
\vspace{-0.6cm}
\caption{
Diagram (b): Full calculation and expansions
($\Mt=175 \GeV$, $p^2=\MW^2$).}
\label{Gamb-b-mh}
\end{figure}

\vspace{-20pt}
\subsection{Sensitivity of the two-loop top-quark corrections to $\De r$
on the Higgs-boson mass}
\begin{figure}[htb]
\hspace{1.8cm}
% GNUPLOT: LaTeX picture with Postscript
\setlength{\unitlength}{0.1bp}
\special{!
%!PS-Adobe-2.0
%%Creator: gnuplot
%%DocumentFonts: Helvetica
%%BoundingBox: 50 50 626 418
%%Pages: (atend)
%%EndComments
/gnudict 40 dict def
gnudict begin
/Color false def
/Solid false def
/gnulinewidth 5.000 def
/vshift -33 def
/dl {10 mul} def
/hpt 31.5 def
/vpt 31.5 def
/M {moveto} bind def
/L {lineto} bind def
/R {rmoveto} bind def
/V {rlineto} bind def
/vpt2 vpt 2 mul def
/hpt2 hpt 2 mul def
/Lshow { currentpoint stroke M
  0 vshift R show } def
/Rshow { currentpoint stroke M
  dup stringwidth pop neg vshift R show } def
/Cshow { currentpoint stroke M
  dup stringwidth pop -2 div vshift R show } def
/DL { Color {setrgbcolor Solid {pop []} if 0 setdash }
 {pop pop pop Solid {pop []} if 0 setdash} ifelse } def
/BL { stroke gnulinewidth 2 mul setlinewidth } def
/AL { stroke gnulinewidth 2 div setlinewidth } def
/PL { stroke gnulinewidth setlinewidth } def
/LTb { BL [] 0 0 0 DL } def
/LTa { AL [1 dl 2 dl] 0 setdash 0 0 0 setrgbcolor } def
/LT0 { PL [] 0 1 0 DL } def
/LT1 { PL [4 dl 2 dl] 0 0 1 DL } def
/LT2 { PL [2 dl 3 dl] 1 0 0 DL } def
/LT3 { PL [1 dl 1.5 dl] 1 0 1 DL } def
/LT4 { PL [5 dl 2 dl 1 dl 2 dl] 0 1 1 DL } def
/LT5 { PL [4 dl 3 dl 1 dl 3 dl] 1 1 0 DL } def
/LT6 { PL [2 dl 2 dl 2 dl 4 dl] 0 0 0 DL } def
/LT7 { PL [2 dl 2 dl 2 dl 2 dl 2 dl 4 dl] 1 0.3 0 DL } def
/LT8 { PL [2 dl 2 dl 2 dl 2 dl 2 dl 2 dl 2 dl 4 dl] 0.5 0.5 0.5 DL } def
/P { stroke [] 0 setdash
  currentlinewidth 2 div sub M
  0 currentlinewidth V stroke } def
/D { stroke [] 0 setdash 2 copy vpt add M
  hpt neg vpt neg V hpt vpt neg V
  hpt vpt V hpt neg vpt V closepath stroke
  P } def
/A { stroke [] 0 setdash vpt sub M 0 vpt2 V
  currentpoint stroke M
  hpt neg vpt neg R hpt2 0 V stroke
  } def
/B { stroke [] 0 setdash 2 copy exch hpt sub exch vpt add M
  0 vpt2 neg V hpt2 0 V 0 vpt2 V
  hpt2 neg 0 V closepath stroke
  P } def
/C { stroke [] 0 setdash exch hpt sub exch vpt add M
  hpt2 vpt2 neg V currentpoint stroke M
  hpt2 neg 0 R hpt2 vpt2 V stroke } def
/T { stroke [] 0 setdash 2 copy vpt 1.12 mul add M
  hpt neg vpt -1.62 mul V
  hpt 2 mul 0 V
  hpt neg vpt 1.62 mul V closepath stroke
  P  } def
/S { 2 copy A C} def
end
}
\begin{picture}(2880,1576)(0,0)
\special{"
gnudict begin
gsave
50 50 translate
0.100 0.100 scale
0 setgray
/Helvetica findfont 100 scalefont setfont
newpath
-500.000000 -500.000000 translate
LTa
600 251 M
0 1274 V
LTb
600 349 M
63 0 V
2034 0 R
-63 0 V
600 545 M
63 0 V
2034 0 R
-63 0 V
600 741 M
63 0 V
2034 0 R
-63 0 V
600 937 M
63 0 V
2034 0 R
-63 0 V
600 1133 M
63 0 V
2034 0 R
-63 0 V
600 1329 M
63 0 V
2034 0 R
-63 0 V
600 1525 M
63 0 V
2034 0 R
-63 0 V
600 251 M
0 63 V
0 1211 R
0 -63 V
810 251 M
0 63 V
0 1211 R
0 -63 V
1019 251 M
0 63 V
0 1211 R
0 -63 V
1229 251 M
0 63 V
0 1211 R
0 -63 V
1439 251 M
0 63 V
0 1211 R
0 -63 V
1649 251 M
0 63 V
0 1211 R
0 -63 V
1858 251 M
0 63 V
0 1211 R
0 -63 V
2068 251 M
0 63 V
0 1211 R
0 -63 V
2278 251 M
0 63 V
0 1211 R
0 -63 V
2487 251 M
0 63 V
0 1211 R
0 -63 V
2697 251 M
0 63 V
0 1211 R
0 -63 V
600 251 M
2097 0 V
0 1274 V
-2097 0 V
600 251 L
LT0
1499 830 M
180 0 V
726 1525 M
42 -72 V
42 -60 V
42 -50 V
42 -44 V
43 -39 V
42 -34 V
42 -32 V
42 -30 V
42 -27 V
42 -25 V
42 -24 V
43 -22 V
42 -20 V
42 -18 V
42 -15 V
42 -15 V
42 -18 V
43 -19 V
42 -20 V
42 -20 V
42 -21 V
42 -21 V
42 -22 V
42 -22 V
43 -21 V
42 -22 V
42 -22 V
42 -21 V
42 -22 V
42 -21 V
42 -22 V
43 -21 V
42 -21 V
42 -21 V
42 -21 V
42 -21 V
42 -21 V
43 -21 V
42 -20 V
42 -20 V
42 -21 V
42 -20 V
42 -20 V
42 -20 V
43 -20 V
42 -19 V
LT1
1499 720 M
180 0 V
726 1525 M
42 -52 V
42 -44 V
42 -39 V
42 -34 V
43 -30 V
42 -27 V
42 -25 V
42 -23 V
42 -20 V
42 -17 V
42 -17 V
43 -18 V
42 -18 V
42 -19 V
42 -18 V
42 -18 V
42 -19 V
43 -17 V
42 -18 V
42 -17 V
42 -17 V
42 -16 V
42 -17 V
42 -16 V
43 -15 V
42 -15 V
42 -16 V
42 -14 V
42 -15 V
42 -14 V
42 -14 V
43 -14 V
42 -14 V
42 -13 V
42 -13 V
42 -13 V
42 -13 V
43 -13 V
42 -12 V
42 -12 V
42 -13 V
42 -12 V
42 -11 V
42 -12 V
43 -12 V
42 -11 V
LT2
1499 610 M
180 0 V
726 1525 M
42 -65 V
42 -55 V
42 -47 V
42 -42 V
43 -37 V
42 -33 V
42 -31 V
42 -28 V
42 -26 V
42 -24 V
42 -22 V
43 -19 V
42 -17 V
42 -18 V
42 -20 V
42 -20 V
42 -21 V
43 -21 V
42 -20 V
42 -21 V
42 -21 V
42 -20 V
42 -21 V
42 -20 V
43 -19 V
42 -20 V
42 -19 V
42 -20 V
42 -19 V
42 -18 V
42 -19 V
43 -18 V
42 -18 V
42 -18 V
42 -17 V
42 -18 V
42 -17 V
43 -17 V
42 -17 V
42 -17 V
42 -16 V
42 -17 V
42 -16 V
42 -16 V
43 -16 V
42 -16 V
LT3
1499 500 M
180 0 V
726 1525 M
42 -71 V
42 -56 V
42 -46 V
42 -39 V
43 -34 V
42 -29 V
42 -27 V
42 -24 V
42 -23 V
42 -20 V
42 -20 V
43 -18 V
42 -18 V
42 -16 V
42 -15 V
42 -14 V
42 -11 V
43 -9 V
42 -12 V
42 -14 V
42 -16 V
42 -17 V
42 -18 V
42 -20 V
43 -19 V
42 -21 V
42 -20 V
42 -21 V
42 -22 V
42 -21 V
42 -22 V
43 -22 V
42 -22 V
42 -22 V
42 -23 V
42 -22 V
42 -22 V
43 -23 V
42 -22 V
42 -23 V
42 -22 V
42 -23 V
42 -22 V
42 -23 V
43 -22 V
42 -23 V
LT4
1499 390 M
180 0 V
726 1525 M
42 -59 V
42 -43 V
42 -32 V
42 -25 V
43 -19 V
42 -16 V
42 -14 V
42 -11 V
42 -10 V
42 -9 V
42 -9 V
43 -8 V
42 -7 V
42 -7 V
42 -7 V
42 -7 V
42 -6 V
43 -5 V
42 -4 V
42 -1 V
42 -2 V
42 -6 V
42 -8 V
42 -10 V
43 -12 V
42 -14 V
42 -15 V
42 -16 V
42 -17 V
42 -17 V
42 -18 V
43 -19 V
42 -20 V
42 -20 V
42 -20 V
42 -21 V
42 -22 V
43 -21 V
42 -22 V
42 -22 V
42 -22 V
42 -23 V
42 -23 V
42 -23 V
43 -23 V
42 -24 V
stroke
grestore
end
showpage
}
\put(1439,390){\makebox(0,0)[r]{$\Mt = 225 \GeV$}}
\put(1439,500){\makebox(0,0)[r]{$\Mt = 200 \GeV$}}
\put(1439,610){\makebox(0,0)[r]{$\Mt = 150 \GeV$}}
\put(1439,720){\makebox(0,0)[r]{$\Mt = 125 \GeV$}}
\put(1439,830){\makebox(0,0)[r]{$\Mt = 175 \GeV$}}
\put(2968,151){\makebox(0,0){$\frac{\MH}{\GeV}$}}
\put(-20,888){%
%\special{ps: gsave currentpoint currentpoint translate
%270 rotate neg exch neg exch translate}%
\makebox(0,0)[b]{\shortstack{$\Delta r_{(2)}$}}%
%\special{ps: currentpoint grestore moveto}%
}
\put(2697,151){\makebox(0,0){1000}}
\put(2487,151){\makebox(0,0){900}}
\put(2278,151){\makebox(0,0){800}}
\put(2068,151){\makebox(0,0){700}}
\put(1858,151){\makebox(0,0){600}}
\put(1649,151){\makebox(0,0){500}}
\put(1439,151){\makebox(0,0){400}}
\put(1229,151){\makebox(0,0){300}}
\put(1019,151){\makebox(0,0){200}}
\put(810,151){\makebox(0,0){100}}
\put(600,151){\makebox(0,0){0}}
\put(540,1525){\makebox(0,0)[r]{0}}
\put(540,1329){\makebox(0,0)[r]{-0.0002}}
\put(540,1133){\makebox(0,0)[r]{-0.0004}}
\put(540,937){\makebox(0,0)[r]{-0.0006}}
\put(540,741){\makebox(0,0)[r]{-0.0008}}
\put(540,545){\makebox(0,0)[r]{-0.001}}
\put(540,349){\makebox(0,0)[r]{-0.0012}}
\end{picture}
\vspace{-0.6cm}
\caption{
Two-loop top-quark contributions to 
$\Delta r$ subtracted at $\MH=60\,\GeV$.}
\label{fig:delr1}
\end{figure}
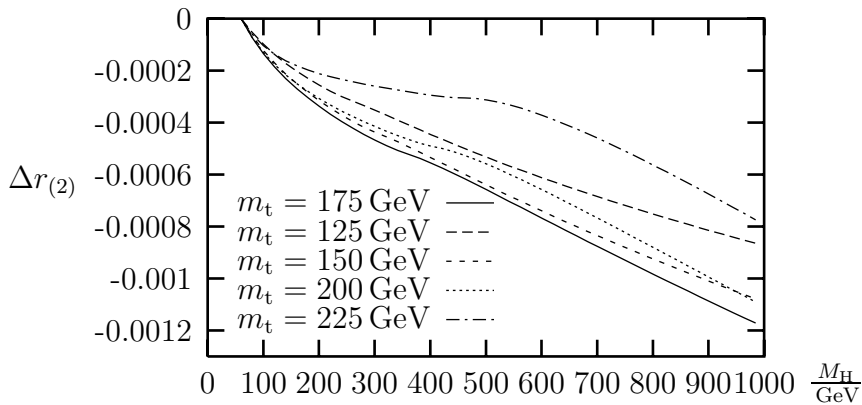

\vspace{-20pt}
\label{sect:DeltaR}
\noindent
In this section we study the sensitivity of the two-loop top-quark 
corrections to $\De r$ to the Higgs-boson mass by considering the quantity
\vspace{-10pt}
\beq
\De r^{\mathrm top}_{(2), {\mathrm subtr}}(\MH) = 
\De r^{\mathrm top}_{(2)}(\MH) - \De r^{\mathrm top}_{(2)}(\MH = 60 \GeV),
\eeq
\vspace{-20pt}
where $\De r^{\mathrm top}_{(2)}(\MH)$ denotes the complete two-loop 
top-quark contribution to $\De r$. 
In \reffi{fig:delr1} 
the variation of $\De r^{\mathrm top}_{(2), {\mathrm subtr}}(\MH)$
with the Higgs-boson mass is shown in the interval $60 \GeV \leq
\MH \leq 1 \TeV$ for various values of $\Mt$. 
The change in $\De r^{\mathrm top}_{(2)}(\MH)$ 
induced by varying $\MH$ in this interval is found to be about
0.001. It is interesting to note that the absolute value of
the correction is maximal just in the region of $\Mt = 175 \GeV$,
i.e.\ for the physical value of the top-quark mass. 

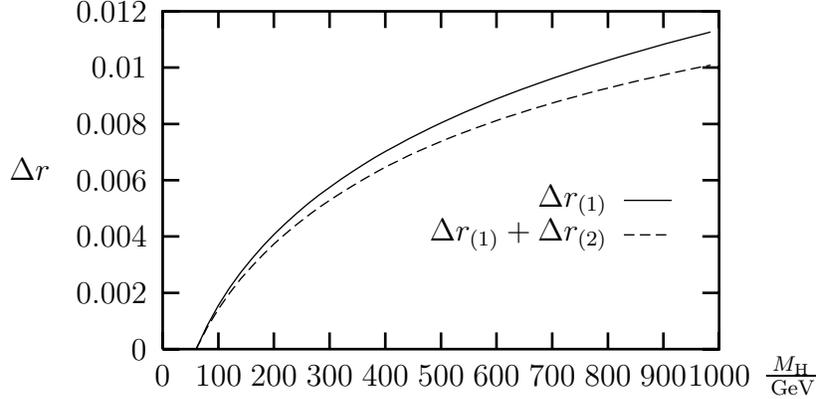
\begin{figure}[htb]
\hspace{1.8cm}
% GNUPLOT: LaTeX picture with Postscript
\setlength{\unitlength}{0.1bp}
\special{!
%!PS-Adobe-2.0
%%Creator: gnuplot
%%DocumentFonts: Helvetica
%%BoundingBox: 50 50 626 418
%%Pages: (atend)
%%EndComments
/gnudict 40 dict def
gnudict begin
/Color false def
/Solid false def
/gnulinewidth 5.000 def
/vshift -33 def
/dl {10 mul} def
/hpt 31.5 def
/vpt 31.5 def
/M {moveto} bind def
/L {lineto} bind def
/R {rmoveto} bind def
/V {rlineto} bind def
/vpt2 vpt 2 mul def
/hpt2 hpt 2 mul def
/Lshow { currentpoint stroke M
  0 vshift R show } def
/Rshow { currentpoint stroke M
  dup stringwidth pop neg vshift R show } def
/Cshow { currentpoint stroke M
  dup stringwidth pop -2 div vshift R show } def
/DL { Color {setrgbcolor Solid {pop []} if 0 setdash }
 {pop pop pop Solid {pop []} if 0 setdash} ifelse } def
/BL { stroke gnulinewidth 2 mul setlinewidth } def
/AL { stroke gnulinewidth 2 div setlinewidth } def
/PL { stroke gnulinewidth setlinewidth } def
/LTb { BL [] 0 0 0 DL } def
/LTa { AL [1 dl 2 dl] 0 setdash 0 0 0 setrgbcolor } def
/LT0 { PL [] 0 1 0 DL } def
/LT1 { PL [4 dl 2 dl] 0 0 1 DL } def
/LT2 { PL [2 dl 3 dl] 1 0 0 DL } def
/LT3 { PL [1 dl 1.5 dl] 1 0 1 DL } def
/LT4 { PL [5 dl 2 dl 1 dl 2 dl] 0 1 1 DL } def
/LT5 { PL [4 dl 3 dl 1 dl 3 dl] 1 1 0 DL } def
/LT6 { PL [2 dl 2 dl 2 dl 4 dl] 0 0 0 DL } def
/LT7 { PL [2 dl 2 dl 2 dl 2 dl 2 dl 4 dl] 1 0.3 0 DL } def
/LT8 { PL [2 dl 2 dl 2 dl 2 dl 2 dl 2 dl 2 dl 4 dl] 0.5 0.5 0.5 DL } def
/P { stroke [] 0 setdash
  currentlinewidth 2 div sub M
  0 currentlinewidth V stroke } def
/D { stroke [] 0 setdash 2 copy vpt add M
  hpt neg vpt neg V hpt vpt neg V
  hpt vpt V hpt neg vpt V closepath stroke
  P } def
/A { stroke [] 0 setdash vpt sub M 0 vpt2 V
  currentpoint stroke M
  hpt neg vpt neg R hpt2 0 V stroke
  } def
/B { stroke [] 0 setdash 2 copy exch hpt sub exch vpt add M
  0 vpt2 neg V hpt2 0 V 0 vpt2 V
  hpt2 neg 0 V closepath stroke
  P } def
/C { stroke [] 0 setdash exch hpt sub exch vpt add M
  hpt2 vpt2 neg V currentpoint stroke M
  hpt2 neg 0 R hpt2 vpt2 V stroke } def
/T { stroke [] 0 setdash 2 copy vpt 1.12 mul add M
  hpt neg vpt -1.62 mul V
  hpt 2 mul 0 V
  hpt neg vpt 1.62 mul V closepath stroke
  P  } def
/S { 2 copy A C} def
end
}
\begin{picture}(2880,1576)(0,0)
\special{"
gnudict begin
gsave
50 50 translate
0.100 0.100 scale
0 setgray
/Helvetica findfont 100 scalefont setfont
newpath
-500.000000 -500.000000 translate
LTa
600 251 M
2097 0 V
600 251 M
0 1274 V
LTb
600 251 M
63 0 V
2034 0 R
-63 0 V
600 463 M
63 0 V
2034 0 R
-63 0 V
600 676 M
63 0 V
2034 0 R
-63 0 V
600 888 M
63 0 V
2034 0 R
-63 0 V
600 1100 M
63 0 V
2034 0 R
-63 0 V
600 1313 M
63 0 V
2034 0 R
-63 0 V
600 1525 M
63 0 V
2034 0 R
-63 0 V
600 251 M
0 63 V
0 1211 R
0 -63 V
810 251 M
0 63 V
0 1211 R
0 -63 V
1019 251 M
0 63 V
0 1211 R
0 -63 V
1229 251 M
0 63 V
0 1211 R
0 -63 V
1439 251 M
0 63 V
0 1211 R
0 -63 V
1649 251 M
0 63 V
0 1211 R
0 -63 V
1858 251 M
0 63 V
0 1211 R
0 -63 V
2068 251 M
0 63 V
0 1211 R
0 -63 V
2278 251 M
0 63 V
0 1211 R
0 -63 V
2487 251 M
0 63 V
0 1211 R
0 -63 V
2697 251 M
0 63 V
0 1211 R
0 -63 V
600 251 M
2097 0 V
0 1274 V
-2097 0 V
600 251 L
LT0
2338 812 M
180 0 V
726 251 M
42 89 V
42 76 V
42 66 V
42 58 V
43 53 V
42 48 V
42 44 V
42 40 V
42 38 V
42 36 V
42 33 V
43 31 V
42 30 V
42 28 V
42 27 V
42 26 V
42 24 V
43 23 V
42 23 V
42 21 V
42 21 V
42 20 V
42 19 V
42 19 V
43 18 V
42 18 V
42 17 V
42 16 V
42 16 V
42 15 V
42 15 V
43 15 V
42 14 V
42 14 V
42 14 V
42 13 V
42 13 V
43 13 V
42 12 V
42 12 V
42 12 V
42 12 V
42 11 V
42 11 V
43 11 V
42 11 V
LT1
2338 682 M
180 0 V
726 251 M
42 82 V
42 69 V
42 60 V
42 54 V
43 48 V
42 44 V
42 41 V
42 37 V
42 35 V
42 33 V
42 31 V
43 29 V
42 27 V
42 26 V
42 25 V
42 24 V
42 23 V
43 21 V
42 21 V
42 19 V
42 18 V
42 18 V
42 17 V
42 16 V
43 16 V
42 15 V
42 15 V
42 14 V
42 13 V
42 14 V
42 12 V
43 13 V
42 12 V
42 11 V
42 12 V
42 11 V
42 10 V
43 11 V
42 10 V
42 10 V
42 9 V
42 10 V
42 9 V
42 9 V
43 9 V
42 8 V
stroke
grestore
end
showpage
}
\put(2278,682){\makebox(0,0)[r]{$\Delta r_{(1)}+ \Delta r_{(2)}$}}
\put(2278,812){\makebox(0,0)[r]{$\Delta r_{(1)}$}}
\put(2968,151){\makebox(0,0){$\frac{\MH}{\GeV}$}}
\put(100,888){%
%\special{ps: gsave currentpoint currentpoint translate
%270 rotate neg exch neg exch translate}%
\makebox(0,0)[b]{\shortstack{$\Delta r$}}%
%\special{ps: currentpoint grestore moveto}%
}
\put(2697,151){\makebox(0,0){1000}}
\put(2487,151){\makebox(0,0){900}}
\put(2278,151){\makebox(0,0){800}}
\put(2068,151){\makebox(0,0){700}}
\put(1858,151){\makebox(0,0){600}}
\put(1649,151){\makebox(0,0){500}}
\put(1439,151){\makebox(0,0){400}}
\put(1229,151){\makebox(0,0){300}}
\put(1019,151){\makebox(0,0){200}}
\put(810,151){\makebox(0,0){100}}
\put(600,151){\makebox(0,0){0}}
\put(540,1525){\makebox(0,0)[r]{0.012}}
\put(540,1313){\makebox(0,0)[r]{0.01}}
\put(540,1100){\makebox(0,0)[r]{0.008}}
\put(540,888){\makebox(0,0)[r]{0.006}}
\put(540,676){\makebox(0,0)[r]{0.004}}
\put(540,463){\makebox(0,0)[r]{0.002}}
\put(540,251){\makebox(0,0)[r]{0}}
\end{picture}
\vspace{-0.6cm}
\caption{
One-loop and two-loop top-quark %contribution
contrib.\
to $\Delta r$ subtracted at $\MH=60\,\GeV$.}
\label{fig:delr2}
\end{figure}

\vspace{-8pt}
In \reffi{fig:delr2}
the corresponding one-loop correction, 
\vspace{-10pt}
\beq
\De r_{(1), {\mathrm subtr}}(\MH) =
\De r_{(1)}(\MH) - \De r_{(1)}(\MH= 60 \GeV)
\eeq
\vspace{-20pt}
is shown together with
the combined contribution $\De r_{(1), {\mathrm subtr}}(\MH) + 
\De r^{\mathrm top}_{(2), {\mathrm subtr}}(\MH)$.
As one can
see in the plot, the one-loop and two-loop corrections enter with
different sign, i.e.\ the sensitivity of $\De r$ to the Higgs-boson
mass is lowered by the inclusion of the two-loop top-quark
corrections. The size of the two-loop correction 
$\De r^{\mathrm top}_{(2), {\mathrm subtr}}(\MH)$ is about 10 percent
of the one-loop contribution.

\vspace{-30pt}
\section{Conclusions}
\vspace{-20pt}
\noindent
In this article we have outlined techniques for the calculation of 
electroweak two-loop corrections associated with the top quark or the 
Higgs boson to precision observables like $\De r$. No expansion in the 
particle masses is performed, i.e.\ the calculations are valid for 
arbitrary values of $\Mt$, $\MH$ and the gauge-boson masses. 
The calculations are performed with the help of computer-algebra
packages, which provide a high degree of automatization, and very
efficient C-routines for the numerics, which are based on
one-dimensional integral representations of the two-loop
scalar integrals. For an example we have compared our results to those
obtained via an expansion in the top-quark mass up to next-to-leading
order. We have furthermore
calculated the sensitivity of the two-loop top-quark
corrections to $\De r$ with respect to the Higgs-boson mass and compared
these contributions to the one-loop result.

\vspace{-30pt}
\begin{ack}
\vspace{-20pt}
\noindent
We thank B.~Tausk for collaboration at the early stages of this work, and 
P.~Gambino for sending us his results used for the comparison in Sect.~5.1.
We also thank M.~B\"ohm, G.~Degrassi, P.~Gambino, W.~Hollik, B.~Tausk and
R.~Scharf for useful discussions.
\end{ack}

\end{document}